\documentstyle[multicol,aps,prl,epsf]{revtex}
\begin{document}
\author{Yu-xi Liu, C. P. Sun, and S. X. Yu}
\address{Institute of Theoretical Physics, Academia Sinica, \\
P.O.Box 2735, Beijing 100080, China}
\title{Quantum decoherence of excitons in a leaky
cavity with quasimodes} \maketitle

\begin{abstract}
For the excitons in the quantum well placed within a leaky cavity, the
quantum decoherence of a mesoscopically superposed states is
investigated based on the factorization theory for quantum
dissipation. It is found that the coherence of the exciton
superposition states will decrease in an oscillating form when
the cavity field interacting with the exciton is of the form of
quasimode. The effect of the thermal cavity fields on the
quantum decoherence of the superposition states of the exciton is studied and
it is observed that the  higher the temperature of the environment
is, the  shorter the decoherence characteristic time is.
\end{abstract}


\pagenumbering{roma}

\thispagestyle{empty} \vspace{15mm} \widetext

\widetext
\vspace{4mm}

\pagenumbering{arabic}
\begin{multicols}{2}
\section{\ Introduction}
The quantum coherence of superposition states is a basic
principle governing the quantum world. The question  why
macroscopic superposition states are not observable has been
raised by Schr\"odinger in his famous cat paradox~\cite{zur,Har}.
Recent experiments demonstrated that the coherent superposition and its losing
process can be observed in the laboratory, at least in the
mesoscopic domain. In an experiment performed by
Wineland's group~\cite{A}, a superposition of two different coherent
states for an ion oscillating in a harmonic potential was
created as the Schr\"odinger cat. In another experiment~\cite{B},
two coherent states of a cavity mode are also superposed
coherently by the atoms passing the cavity with large
detuning.

In fact, the ideal coherent superposition state
can only live in quantum world which are free from external
influence. A real quantum system  can rarely be completely
isolated from its surrounding environment and is usually coupled
to the external world (also called ``heat bath'') with a large
number of  degrees of freedom. Not only does the usual coupling
allow for an exchange of energy between system and bath causing
dissipation~\cite{leg,yu-sun},  there also exist different
interactions leading to the so-called pure decoherence, the decoherence without
energy dissipation.  So the
quantum mechanical interference effects are destroyed very rapidly
due to the two influences of the environment, quantum dissipation
and decoherence. When they happen, a superposition state of the
system evolves into a statistical mixture state.

In passing years,
quantum decoherence has been experimentally studied for various
systems by using different techniques, such as the ``which-way"
experiments using atoms in Bragg scattering~\cite{wwe1} and the
Aharonov-Bohm electrons in mesoscopic system~\cite{wwe2}.
Recently, the decoherence of superposed motional states of a
trapped single atom, which is induced by coupling the atom or ion to engineered
reservoirs, was tested quantitatively in the experiments with
cooling atoms and ion~\cite{cj}.

This paper is devoted to studying quantum decoherence in a
solid-state-system, specially in the system of excitons. The
motivation to investigate such kind of systems is to consider the practical
realization of the quantum information processes, such as quantum
computing and quantum communications ~\cite{qi}. The coherence is
the essential requirement for the quantum information, but  the
decoherence  will result in  errors in the process of the
computation and quantum communication. So the quantum decoherence
is the biggest obstacle for its implement~\cite{szl}. To
overcome quantum decoherence, one should know its dynamic details
theoretically and experimentally in various physical systems that
are the potential carriers of quantum information. In solid-state
system, maximally entangled states of Bell-type for exciton in
two coupled quantum dots, and Greenberge-Horre-Zeilinger type for
three coupled dots have been investigated~\cite{C}. The decoherence
effects on the generation of entangled
states of exciton have been investigated for the
coupled quantum dot systems~\cite{D}.

We will touch upon this problem for the system of  exciton in the
quantum well immersed in a leaky cavity with the dissipative
cavity fields described by quasi-modes~\cite{scu}.

In the case of low excitation, the collective behavior of
many-molecule can be described by a bosonic exciton~\cite{Hak}.
Therefore the total system can be modeled as a standard harmonic
oscillator-bath (or environment) system. The coupling strength
between the cavity fields and the excitons obeys the Lorentz
spectral distribution around $\Omega$. For such a typical system
without any practical concerning, one of the authors (CPS) and his
collaborators even systematically studied its quantum dissipation
and decoherence in association with quantum measurement problem by
consider the factorization \cite{yu-sun,s1,s2,s3,s4} and partial
factorization \cite{s5} of the wave function of the total system.
This paper will generalize the partial factorizaton method to
discuss quantum decoherence of excitons in a leaky cavity.

The paper is organized as follows. In section II, we first give a
theoretical model and do our best to find  an analytical solutions
of the Heisenberg operators of the cavity fields and the exciton.
In section III, by investigating the decoherence of the
superposition state of the exciton, we find that, because of the
effects of the environment on the system of exciton, the coherence
of the superposition states of the exciton is suppressed in a
oscillating-decaying way. In section IV, we study the quantum
decoherence with the environment of thermal fields at finite
temperature. Finally we give our conclusions and some necessary
remarks.

\section{Theoretical Model and exact solutions}
We consider  a quantum well placed within  a leaky Fabry-Perot
cavity~\cite{scu}. The quantum well lies in the center of the
cavity. It has a ideal cubic lattice with $N$ lattice sites. It is
so thin that it only contains one molecular layer. We assume that
$N$ identical two-level molecules distribute into these lattice
sites. All these molecules have equivalent mode positions, so they
have the same coupling constant. The density of excitation for the
molecules is somewhat low and the inter-molecular interactions are
neglected. It is also assumed that the direction of the dipole
moment for molecules and  wave vectors of the cavity fields  are
perpendicular to the surface of the quantum well. These molecules
interact resonantly with a quasi-mode field with the frequency
$\Omega$. By using Dicke model~\cite{Dik}, we write down the
Hamiltonian for the quantum well and the cavity fields under the
rotating wave approximation:

\begin{equation}
H=\hbar \Omega
S_z+\hbar\sum_{j}\omega_{j}\hat{a}^{\dagger}_{j}\hat{a}_j +\hbar
\sum_{j} g_{j}
(\hat{a}^{\dagger}_{j}S_{+}+\hat{a}_{j}^{\dagger}S_{-})
\end{equation}
with the collective operators
\begin{equation}
S_Z=\sum_{n=1}^Ns_z(n),\quad S_{\pm}=\sum_{n=1}^Ns_{\pm}(n),
\end{equation}
where $s_z(n)=\frac 12(|e_n\rangle\langle e_n|-|g_n\rangle\langle
g_n|)$,
$s_{+}(n)=|e_n\rangle\langle g_n|$ and $%
s_{-}(n)=|g_n\rangle\langle e_n|$ are quasi-spin operators of the
n-th molecule. Here $|e_n\rangle$ and $|g_n\rangle$ the excited
state and the ground state of n-th molecule, $\Omega$ is a
transition frequency of the isolate molecule. Operators
$\hat{a}_{j}^{\dagger}(\hat{a}_{j})$ are  creation (annihilation)
operators of the field modes which labeled by continuous index $j$
and the field frequency of each mode is denoted by $\omega_{j}$.
The coupling constant $g_{j}$ between the molecules and the cavity
fields takes a simple form which is proportional to a Lorentzian
\begin{equation}
g_{j}=\frac{\eta\Gamma}{\sqrt{(\omega_{j}-\Omega)^{2}+\Gamma^{2}}},
\end{equation}
where $\eta$ depends on the atomic dipole~\cite{Law} and $\Gamma$ is
the decay rate of a quasi-mode of the cavity with a frequency
$\Omega$. In this paper we restrict our investigation to only one
quasi-mode cavity. Because the excitation of the molecules is
somewhat low, so we will make a bosonic approximation~\cite{Hak,l,c}
$\hat{b}=\frac{S}{\sqrt{N}}$ and $\hat{b}^{\dagger}
=\frac{{S}^{\dagger}}{\sqrt{N}}$ with $[\hat{b},
\hat{b}^{\dagger}]=1$. Then the interaction between the cavity
field and the quantum  well occurs via excitons.  The Hamiltonian
(1) is changed into
\begin{equation}
H=\hbar \Omega \hat{b}^{\dagger}\hat{b}+
\sum \hbar \omega_{j} \hat{a}_{j}^{\dagger}\hat{a}_{j}
+\sum g(\omega_{j})(\hat{a}_{j}^{\dagger}\hat{b}+
\hat{b}^{\dagger}\hat{a}_{j})
\end{equation}
with $g(\omega_{j})=\sqrt{N}g_{j}$.
In the terms of
Hamiltonian (4), we may write the Heisenberg equations of motion for the
operators of the field modes $\hat{a}_{k}(\hat{a}^{\dagger}_{k})$ and
the excitons $\hat{b}^{\dagger}(\hat{b})$:
\begin{mathletters}
\begin{eqnarray}
\frac{\partial{\hat{b}}}{\partial t}&=&-i\Omega \hat{b}-i\sum g(\omega_{j})
\hat{a}_{j},\\
\frac{\partial{\hat{a}_{j}}}{\partial t}&=&-i\omega_{j} \hat{a}_{j}
-i g(\omega_{j}) \hat{b}.
\end{eqnarray}
\end{mathletters}
The integral equation of Eq.(5b) may be written as
\begin{equation}
\hat{a}_{j}(t)=\hat{a}_{j}(0)e^{-i\omega t}-ig(\omega_{j})\int_{0}^{t}
\hat{b}(t^{\prime})e^{-i\omega_{j}(t-t^{\prime})}{\rm d}t^{\prime}.
\end{equation}
We firstly substitute Eq.(6) into Eq.(5a) and  eliminate the field operators.
Secondly, we let $\hat{b}(t)=\hat{B}(t)e^{-i\Omega t}$ to remove the
high-frequency behavior from Eq.(5a). Then the equation of motion for the
slowly varied exciton operator  is
\begin{equation}
\frac{\partial{\hat{B}}}{\partial t}=-\int^{t}_{0}
\hat{B}(t^{\prime})K(t-t^{\prime}) {\rm d}t^{\prime}
+\eta(t)
\end{equation}
with the general memory kernal function
$K(t-t^{\prime})=\sum_{j}|g(\omega_{j})|^{2}
e^{-i(\omega_{j}-\Omega)(t-t^{\prime})}$ and
$\eta(t)=-i\sum_{j}g(\omega_{j})\hat{a}_{j}(0)e^{-i(\omega_{j}-\Omega)t}$.
By use of the Laplace transformation, we could solve $\hat{B}(t)$ and  find
$\hat{b}(t)$
\begin{mathletters}
\begin{equation}
\hat{b}(t)=\hat{B}(t)e^{-i\Omega t}=
[u(t)\hat{b}(0)+\sum u_{j}(t)\hat{a}_{j}(0)]e^{-i\Omega t}
\end {equation}
where the time-dependent coefficients are
\begin{eqnarray}
u(t)&=&{\cal L}^{-1}\left \{ \tilde{u}(p) \right \} \\
\tilde{u}(p)&=& \left \{p+\tilde{K}(p) \right \}^{-1}
\end{eqnarray}
and
\begin{equation}
u_{j}(t)={\cal L}^{-1} \left \{  \frac{g(\omega_{j})}{p+i(\omega_{j}-\Omega)}
\tilde{u}(p) \right \}
\end{equation}
\end{mathletters}
${\cal L}^{-1}$ denotes the inverse Laplace transformation and $\tilde{K}(p)$ is
the Laplace transformation of the general memory kernal function $K(t-t^{\prime})$.
Substituting Eqs.(8) into Eq.(6), we have
\begin{equation}
\hat{a}_{j}(t)=e^{-i\omega_{j} t}\hat{a}_{j}(0)
+v_{j}(t)\hat{b}(0)+\sum v_{j,j^{\prime}}(t) \hat{a}_{j^{\prime}}(0)
\end{equation}
here, the time-dependent coefficients $v_{j}(t)$ and $v_{j,j^{\prime}}(t)$
are determined by
\begin{mathletters}
\begin{eqnarray}
&&v_{j}(t)=-ig(\omega_{j})e^{-i\omega t}\int^{t}_{0}u(t^{\prime})
e^{i(\omega_{j}-\Omega)t^{\prime}}{\rm d}t^{\prime},\\
&&v_{j,j^{\prime}}(t)=-g(\omega_{j})e^{-i\omega t}\int_{0}^{t}u_{j}(t^\prime)
e^{i(\omega_{j}-\Omega)t^{\prime}}{\rm d}t^{\prime}.
\end{eqnarray}
\end{mathletters}
In order to obtain  above time-dependent coefficients ,
we begin to solve the function $K(t-t^{\prime})$ by changing the sum $\sum_{j}$
into the integration $\frac{L}{\pi c}\int_{0}^{\infty}{\rm d}\omega_{j}$ which
$L$ is the length of the cavity and $c$ is the speed of the light in the
vacuum~\cite{rl}, that is
\begin{equation}
K(t-t^{\prime})=\frac{\eta^{2}\Gamma^{2}NL}{\pi c}
\int_{0}^{\infty}\frac{e^{-i(\omega_{j}-\Omega)(t-t^{\prime})}}
{(\omega_{j}-\Omega)^{2}+\Gamma^{2}}{\rm d}\omega_{j}.
\end{equation}
If we assume that $\Omega$ is much larger than all other quantities of the
dimension of frequency, and $\Gamma$ is small quantity. We may adopt to
the standard approximation of extending the lower limit of the integral Eq.(11)
to $-\infty$. By integrating eq.(11) we have:
\begin{equation}
K(t-t^{\prime})=M\Gamma  e^{-\Gamma|t-t^{\prime}|}
\end{equation}
with $M=\frac{N\eta^{2}L}{c}$. In the following calculation, we only need
time-dependent coefficients  $u(t)$ and $v_{j}(t)$. So after we give the
Laplace form of the function $K(t-t^{\prime})$, we will obtain $u(t)$
by use of Eqs.(8b-8c) as following
\begin{equation}
u(t)=[cos(\frac{\Theta}{2} t)+\frac{\Gamma}{\Theta}sin(\frac{\Theta}{2} t)]
e^{-\frac{\Gamma}{2}t}
\end{equation}
where ${\Theta}=\sqrt{4M\Gamma-\Gamma^{2}}$. $v_{j}(t)$ also can be obtained
by integral eq.(10a) as

\begin{eqnarray}
v_{j}(t)=-ig(\omega_{j})(\frac{1-i\frac{\Gamma}{\Theta}}{2})
\frac{e^{i(\frac{\Theta}{2}-\Omega)t-\frac{\Gamma}{2}t}-e^{-i\omega_{j}t}}
{i(\frac{\Theta}{2}+\omega_{j}-\Omega)-\frac{\Gamma}{2}} \nonumber \\
-ig(\omega_{j})(\frac{1+i\frac{\Gamma}{\Theta}}{2})
\frac{e^{-i(\frac{\Theta}{2}+\Omega)t-\frac{\Gamma}{2}t}-e^{-i\omega_{j}t}}
{i(\omega_{j}-\frac{\Theta}{2}-\Omega)-\frac{\Gamma}{2}}
\end{eqnarray}

\section{ decoherence of  mesoscopic  superposition states of exciton }
If we prepare a superposition state for the system of the  exciton,
that is, the exciton initially is in the state
 $C|\alpha_{1}\rangle+D|\alpha_{2}\rangle$
where $\alpha_{1}$ or $\alpha_{2}$ is a coherent state for the exciton,
and the cavity fields are in the vacuum states  $\prod_{j}|0\rangle_{j}$
(zero temperature). Thus  the  whole initial state for the exciton
and the  cavity fields is the product  of the initial state of the
exciton and the cavity fields.
\begin{equation}
|\Psi(0)\rangle=(C|\alpha_{1}\rangle+D|\alpha_{2}\rangle)
\bigotimes\prod_{j}|0\rangle _{j}.
\end{equation}
In order to discuss the coherence properties for the system of excitons, we
have to calculate the time evolution of the wave function. For this
purpose, we firstly write a state for the whole system  at any time $t$ by
virtue of the evolution operators of the whole system
$U(t)=e^{-i\frac{H t}{\hbar}}$:
\begin{equation}
|\Psi(t)\rangle=U(t)|\Psi(0)\rangle=U(t)(C|\alpha_{1}\rangle+D|\alpha_{2}\rangle)
\bigotimes \prod_{j}|0\rangle_{j}.
\end{equation}
Because for any coherent state of the exciton we have
\begin{equation}
|\alpha\rangle=exp(\alpha b^{\dagger}(0)-\alpha^{*}b(0))|0\rangle
\end{equation}
So we have
\begin{eqnarray}
|\Psi(t)\rangle&=&U(t)[C e^{\alpha_{1} b^{\dagger}(0)-\alpha_{1}^{*}b(0)}+
D e^{\alpha_{2} b^{\dagger}(0)-\alpha^{*}_{2}b(0)}]|0\rangle \nonumber \\
&&\bigotimes \prod_{j}|0\rangle_{j}
\end{eqnarray}
Considering the role of the evolution operator $U(t)$, that is,
$U^{\dagger}(t)OU(t)=O(t)$  and $U(t)|0\rangle=|0\rangle$ for any operator $O$.
We interpolate operator $U^{\dagger}(t)U(t)$ into Eq.(18) and
use the property of the evolution operator $U(t)$.  we
easily obtain the wave function of the whole system at any time $t$
\begin{eqnarray}
|\Psi(t)\rangle&=&C|\alpha_{1}u^{*}(t)\rangle\bigotimes
\prod_{j}|\alpha_{1}u^{*}_{j}(t)\rangle_{j} \nonumber \\
&+&D|\alpha_{2}u^{*}(t)\rangle\bigotimes
\prod_{j}|\alpha_{2}u^{*}_{j}(t)\rangle_{j}
\end{eqnarray}
We could calculate the reduced density matrix of the exciton system  at
any time $t$ by $Tr_{R}(|\Psi(t)\rangle\langle\Psi(t)|)$, and obtain the
decoherence factor by calculating one of the non-diagonal elements, such as
\begin{eqnarray}
F(t)&=&\prod_{j}\langle\alpha^{*}_{1}u_{j}(t)|\alpha_{2}u^{*}_{j}(t)\rangle\nonumber \\
&=&e^{(-\frac{1}{2}|\alpha_{1}|^{2}-\frac{1}{2}|\alpha_{2}|^{2}+
\alpha^{*}_{1}\alpha_{2})\sum_{j}|u_{j}(t)|^{2}}.
\end{eqnarray}
We know that $[\hat{b}(t), \hat{b}^{\dagger}(t)]=1$. From Eq.(8a) and its
complex conjugate, we have
\begin{eqnarray}
&&\sum_{j}|u_{j}(t)|^{2}=1-|u(t)|^{2} \nonumber \\
&&=1-|cos(\frac{\Theta}{2} t)+\frac{\Gamma}{\Theta}sin(\frac{\Theta}{2} t)|^{2}
e^{-\Gamma t}.
\end{eqnarray}
Eq.(20) becomes into:
\begin{equation}
F(t)=e^{(-\frac{1}{2}|\alpha_{1}|^{2}-\frac{1}{2}|\alpha_{2}|^{2}+
\alpha^{*}_{1}\alpha_{2})(1-|cos(\frac{\Theta}{2} t)+
\frac{\Gamma}{\Theta}sin(\frac{\Theta}{2} t)|^{2}
e^{-\Gamma t})}
\end{equation}
so with the evolution of the time, the coherence of two coherent states of
the excitons is suppressed. It is evident when the time $t \rightarrow \infty$,
 because of the environment effect, the energy
of the exciton will be  dissipated and states of the exciton will turn into
vacuum states.  We consider  a behavior of the short time, that is
$\gamma t,  \frac{\Theta}{2} t\ll 1$, then the
decoherence factor is
\begin{equation}
F(t)=e^{(-\frac{1}{2}|\alpha_{1}|^{2}-\frac{1}{2}|\alpha_{2}|^{2}+
\alpha^{*}_{1}\alpha_{2})\Gamma t}
\end{equation}
The characteristic time $t_{d}$ of the decoherence of the exciton superposition
states is  $[(\frac{1}{2}|\alpha_{1}|^{2}+
\frac{1}{2}|\alpha_{2}|^{2}-\alpha^{*}_{1}\alpha_{2})\Gamma]^{-1}$. The
coherent properties of the exciton states is determined by their initial
superposition states and the decay rate of the quasimode. The smaller the
superposition of $|\alpha_{1}\rangle$ and
$|\alpha_{2}\rangle$ is, The shorter  the decoherence time $t_{d}$ is.

Now we consider a special case in which many people are
interested~\cite{c1,c2,c3,c4}, that is,
the system is initially in the odd or even coherent states of the exciton.
We set $\alpha_{1}=-\alpha_{2}=\alpha$, then
the decoherence factor is
$$ F(t)=e^{-2|\alpha|^{2}(1-|cos(\frac{\Theta}{2} t)+
\frac{\Gamma}{\Theta}sin(\frac{\Theta}{2} t)|^{2}
e^{-\Gamma t})}$$
In fig.1, we give a sketch of the decoherence factor as the function of the
time $t$ in a set of parameters. We find that the coherence of the exciton
system will decrease in the oscillating decay form. It is similar to the
case of the experiment~\cite{cj}

In terms of the definition of the coherent state of the exciton, we know that
$|\alpha|^{2}$ is  the mean number of the excitons.
Hence by adjusting the initial number of the excitons we could completely
determined the decoherence time and. The smaller the mean number of the
exciton is, the longer the decoherence time.

\vskip -2cm
\begin{figure}
\epsfxsize=6cm
\centerline{\epsffile{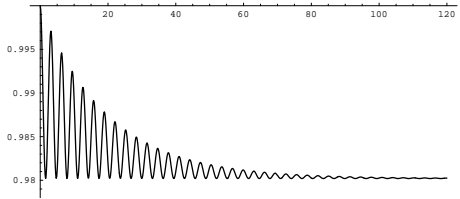}}
\vskip -2cm
\caption[]{$\hbar\Gamma$=0.05 mev, $|\alpha|^{2}$=0.01, $\hbar$M=20 mev }
\end{figure}

\section{temperature effect on coherence}

In this section, we will discuss the temperature effect on the coherent
properties of the exciton superposition states. For a single mode thermal
cavity field of the frequency $\omega_{j}$, its variables are in the thermal
equilibrium mixture of states. Its density operator could be given  using
 Fock states
\begin{equation}
\hat{\rho}_{j}=\frac{1}{\langle n_{j}\rangle+1}
\sum_{n_{j}}(\frac{\langle n_{j}\rangle}{\langle n_{j}\rangle+1})^{n_{j}}
|n_{j}\rangle\langle n_{j}|.
\end{equation}
where the average thermal photon number
$\langle n_{j}\rangle=[exp(\frac{\hbar\omega_{j}}{k_{B}T})-1]^{-1}$.
$k_{B}$ is Boltzman constant. $T$ is temperature of the thermal fields.
under the  P representation eq.(19) is rewritten as
\begin{equation}
\hat{\rho}_{j}=\int \frac{{\rm d}^{2}\alpha_{j}}
{\pi \langle n_{j}\rangle}e^{-\frac{|\alpha_{j}|^{2}}{\langle n_{j}\rangle}}
|\alpha_{j}\rangle\langle\alpha_{j}|
\end{equation}
It is clear that the P-representation of the thermal fields  is given by a
Gaussian distribution. If we assumed that the cavity fields initially are in
the thermal states, then the reduced density operator of the cavity fields is
the multi-mode extension of the thermal field operators, namely,
$\hat{\rho}_{bath}=\prod_{j}\hat{\rho}_{j}$.
The whole density operator are
\begin{equation}
\hat{\rho}=\hat{\rho}_{e}\bigotimes\hat{\rho}_{bath}
\end{equation}
with density operators of exciton
\begin{equation}
\hat{\rho}_{e}=(C|\alpha_{1}\rangle+D|\alpha_{2}\rangle)
(\langle\alpha_{1}|C^{*}+\langle\alpha_{2}|D^{*})
\end{equation}
In order to investigate the effect of the thermal fields on coherence of
the superposition states of the exciton system , we have to calculate the
wave function of the whole system at any time $t$. So we set
\begin{equation}
|\psi(t)\rangle=U(t)(|\alpha\rangle\bigotimes|\alpha_{j}\rangle)
\end{equation}
By using the same method as last section, we could obtain
\begin{eqnarray}
|\psi(t)\rangle&=&|u^{*}(t)\alpha+\sum_{j}u^{*}(t)_{j}\lambda_{j}\rangle \nonumber \\
& &\bigotimes|e^{i\omega_{j}t}\lambda_{j}+v^{*}_{j}(t)\alpha+\sum_{s \not= j}
v^{*}(t)_{s,j}\lambda_{s}\rangle_{j}
\end{eqnarray}
By virtue of normalized condition
\begin{equation}
[\hat{b}(t), \hat{a}^{\dagger}_{j}(t)]=0,
\end{equation}
we could get the decoherence factor by calculating the reduced density matirx
of the exciton system.
\begin{eqnarray}
F(t)&=&Tr_{R}{\hat{\rho}(t)}
=e^{(-\frac{1}{2}|\alpha_{1}|^{2}-\frac{1}{2}|\alpha_{2}|^{2}+\alpha^{*}_{1}
\alpha_{2})(1-|u(t)|^{2}} \times \nonumber  \\
&&\prod_{j}\int{\rm d}^{2}{\lambda_{j}}\frac{1}{\pi\langle n_{j}\rangle}
e^{(-\frac{|\lambda_{j}|^{2}}{\langle n_{j}\rangle}+\zeta\lambda^{*}_{j}
+\zeta^{*}\lambda_{j})}
\end{eqnarray}
with
\begin{equation}
\zeta=\frac{1}{2}(\alpha_{2}-\alpha_{1})
u(t)v_{j}^{*}\lambda^{*}_{j}
\end{equation}
Combining
\begin{equation}
\frac{1}{\pi}\int{\rm d}^{2}\lambda_{j}
 e^{(-\eta |\lambda_{j}|^{2}-\zeta\lambda^{*}_{j}
+\zeta^{*}\lambda)}= \frac{1}{\eta}e^{-\frac{1}{\eta}|\zeta|^{2}}
\hspace{0.2cm} {\rm For} \hspace{0.1cm} {\rm Re}(\eta)\rangle0
\end{equation}
we get the decoherence factor as following:
\begin{mathletters}
\begin{eqnarray}
&&F(t)=exp[(-\frac{1}{2}|\alpha_{1}|^{2}-\frac{1}{2}|\alpha_{2}|^{2}
+\alpha^{*}_{1}\alpha_{2})(1-|u(t)|^{2}] \nonumber \\
&&\times exp[-\frac{1}{4}|\alpha_{1}-\alpha_{2}|^{2}|u(t)|^{2}\beta(t,T)]
\end{eqnarray}
with
\begin{equation}
\beta(t,T)=\sum_{j}
|v_{j}(t)|^{2}\langle n_{j}\rangle
\end{equation}
\end{mathletters}
We may transform $\sum_{j}$ of eq.(34b) into
$\frac{L}{\pi c}\int_{0}^{\infty}{\rm d}\omega$ as reference~\cite{rl}
\begin{equation}
\beta(t, T)= \frac{L}{\pi c}
\int_{0}^{\infty} |v_{j}(t)|^{2}\langle n_{j}\rangle{\rm d}\omega_{j}
\end{equation}

We find that the integration in Eq.(35) is so strongly peaked at the
near $\omega_{j}\approx\Omega$, so we may remove the slowly vary
factor $\langle n_{j}\rangle$. Although we may calculate an exact integration
of Eq.(35) (for the details please see the appendix A), we are more interested
in  the limit of the short time, we finish the integral of Eq.(35)  and obtain
\begin{equation}
\beta(T,t)=\bar{n}{\cal D}\Gamma t
\end{equation}
with $\bar{n}=(e^{\frac{\hbar \Omega}{k_{B}T}}-1)^{-1}$ and ${\cal D}$ is given
in (A6). So in the case of the thermal fields,  we only keep the first order
terms of $\Gamma t$ and $\frac{\Theta}{2} t$,  the decoherence factor of
superposition states of the  exciton system   is
\begin{eqnarray}
F(t)&=&exp[(-\frac{1}{2}|\alpha_{1}|^{2}-\frac{1}{2}|\alpha_{2}|^{2}
+\alpha^{*}_{1}\alpha_{2})\Gamma t] \times \nonumber \\
&\times&exp[-\frac{\bar{n}{\cal D}}{4}|\alpha_{1}-\alpha_{2}|^{2}\Gamma t]
\end{eqnarray}
Under the condition of the high temperature,
The decoherence factor is simplified
\begin{eqnarray}
&&F(t)=exp[(-\frac{1}{2}|\alpha_{1}|^{2}-\frac{1}{2}|\alpha_{2}|^{2}
+\alpha^{*}_{1}\alpha_{2})\Gamma t] \nonumber \\
&&\times exp[-\frac{\cal D}{4}|\alpha_{1}-\alpha_{2}|^{2}
\frac{Tk_{B}}{\hbar\Omega}\Gamma t]
\end{eqnarray}
So the characteristic time
$t_{d}$ of the decoherence of the exciton system is
\begin{equation}
t_{d}=[(\frac{1}{2}|\alpha_{1}|^{2}+\frac{1}{2}|\alpha_{2}|^{2}
-\alpha^{*}_{1}\alpha_{2}+\frac{\cal D}{4}|\alpha_{1}-\alpha_{2}|^{2}
\frac{Tk_{B}}{\hbar\Omega})\Gamma]^{-1}
\end{equation}
It is shown that the higher the temperature of the environment is,
the short the time of decoherence of the exciton system is.
if the exciton system is initial in the odd or even coherent states, then
the characteristic time is
$t_{d}=
[|\alpha_{1}|^{2}(2+\frac{{\cal D}Tk_{B}}{\hbar\Omega})\Gamma]^{-1}$
So the decoherence time becomes more shorter  when the cavity fields are
initially in the thermal radiation.
\begin{section}{conclusions}\end{section}
The decoherence of the mesoscopic superposition states  for the exciton
in a quasimode cavity is investigated. We find that the coherence of the
superposition states of exciton
is reduced by  the interaction between the cavity fields and the excitons. For
long time we find that because of the environment effect, the energy of the
exciton will dissipated and the states of the exciton will turn into turn into
vacuum states.  We find for a short time process, the smaller the superposition
of two coherent states of exciton is, the shorter the decoherence time is.
The temperature effect on coherence of the system of exciton also is studied
by virtue of the P representation.
We find that the more higher of the temperature of the environment is,
the more shorter of the  decoherence time is.

{\bf Acknowledges}
One of authors (Yu-xi Liu) would like to  express his sincere thanks to
Dr. X. X. Yi for his many useful discussions. This work is supported
partially by  the NSF of China and ``9-5 Pandeng '' project.
\end{multicols}
\widetext
\vspace*{30pt} \centerline{\large {\bf Appendix A \hspace{0.2cm} Solution of
$\beta(T,t)$}} \setcounter{equation}{0}

\vspace{1cm} In this appendix, we will give an integration of $\beta(T,t)$
in details. From Eq.(14) and Eq.(35), we have
\begin{eqnarray}
\beta(T,t)&=& \left \{ \bar{n}\frac{L\eta^{2}\Gamma^{2}(\Theta-i\Gamma)^{2}%
} {4\pi \Theta^{2} c}\int^{\infty}_{0} \frac{1+e^{-\frac{\Gamma}{2}t+i\frac{%
\Theta}{2} t}- e^{i(\omega_{j}+\frac{\Theta}{2}-\Omega)t-\frac{\Gamma}{2}t}
-e^{-i(\omega_{j}-\frac{\Theta}{2}-\Omega)t-\frac{\Gamma}{2}t}} {%
[(\omega_{j}-\Omega)^{2}+\Gamma^{2}] [(\omega_{j}+\frac{\Theta}{2}-\Omega)+i%
\frac{\Gamma}{2}] [(\omega_{j}-\frac{\Theta}{2}-\Omega)-i\frac{\Gamma}{2}]}
{\rm d}\omega_{j}+h.c. \right \}  \nonumber \\
&+&\frac{\bar{n}L\eta^{2}\Gamma^{2}[\Theta^{2}+\Gamma^{2}]} {4\pi \Theta^{2}
c} \int_{0}^{\infty}\frac{1+e^{-\Gamma t}- 2e^{-\frac{\Gamma}{2}%
}cos(\omega_{j}+\frac{\Theta}{2}-\Omega)t} {[(\omega_{j}-\Omega)^{2}+%
\Gamma^{2}][(\omega_{j}+\frac{\Theta}{2}-\Omega)^{2} +\frac{\Gamma^{2}}{4}]}%
{\rm d}\omega_{j}  \nonumber \\
&+& \frac{\bar{n}L\eta^{2}\Gamma^{2}[\Theta^{2}+\Gamma^{2}]} {4\pi
\Theta^{2} c} \int_{0}^{\infty} \frac{[1+e^{-\Gamma t}- 2e^{-\frac{\Gamma}{2}%
}cos(\omega_{j}-\frac{\Theta}{2}-\Omega)t]} {[(\omega_{j}-\Omega)^{2}+%
\Gamma^{2}][(\omega_{j}-\frac{\Theta}{2}-\Omega)^{2} +\frac{\Gamma^{2}}{4}]}%
{\rm d}\omega_{j}
\end{eqnarray}
\widetext
We set $x=\omega_{j}-\Omega$, and extend the lower limit of the integral
Eq.(A1) to $-\infty$ as we do in the section {\bf II}, then finish following
some integral formulations
\begin{eqnarray}
&&\int_{-\infty}^{\infty}\frac{{\rm d}x}{[x^{2}+\Gamma^{2}] [x+\frac{\Theta}{%
2}+i\frac{\Gamma}{2}][x-\frac{\Theta}{2}-i\frac{\Gamma}{2}]} = \frac{\pi}{%
\Gamma(\frac{\Theta}{2}+i\frac{3}{2}\Gamma) (i\frac{1}{2}\Gamma-\frac{\Theta%
}{2})}+ \frac{i2\pi}{(2\frac{\Theta}{2}+i\Gamma)(\frac{\Theta}{2}+i\frac{3}{2%
}\Gamma) (-i\frac{1}{2}\Gamma+\frac{\Theta}{2})} \\
&&\int_{-\infty}^{\infty}\frac{e^{i(x+\frac{\Theta}{2})t-\frac{\Gamma}{2}t}%
{\rm d}x}{[x^{2}+\Gamma^{2}] [x+\frac{\Theta}{2}+i\frac{\Gamma}{2}][x-\frac{%
\Theta}{2}-i\frac{\Gamma}{2}]} =\frac{\pi e^{(i\frac{\Theta}{2}-\frac{3}{2}%
\Gamma)t}} {\Gamma(\frac{\Theta}{2}+i\frac{3}{2}\Gamma) (i\frac{1}{2}\Gamma-%
\frac{\Theta}{2})}+ \frac{i2\pi e^{(i2\frac{\Theta}{2}-\Gamma)t}} {(2\frac{%
\Theta}{2}+i\Gamma)(\frac{\Theta}{2}+i\frac{3}{2}\Gamma) (-i\frac{1}{2}%
\Gamma+\frac{\Theta}{2})} \\
&&\int_{-\infty}^{\infty} \frac{cos(x\pm\frac{\Theta}{2})t]} {%
[x^{2}+\Gamma^{2}][(x\pm\frac{\Theta}{2})^{2}+\frac{\Gamma^{2}}{4}]}{\rm d}x
= \frac{\pi e^{-\Gamma t}[(\frac{\Theta}{2}^{2}-\frac{3}{4}\Gamma^{2})cos(%
\frac{\Theta}{2} t)+ 2\Gamma\frac{\Theta}{2} sin(\frac{\Theta}{2} t)]+2\pi
e^{-\frac{\Gamma}{2}t}(\frac{\Theta}{2}^{2}+ \frac{3}{4}\Gamma^{2})}{\Gamma(%
\frac{\Theta}{2}^{2}+\frac{9}{4}\Gamma^{2}) (\frac{\Theta}{2}^{2}+\frac{%
\Gamma^{2}}{4})} \\
&&\int_{-\infty}^{\infty}\frac{{\rm d}x}{[x^{2}+\Gamma^{2}] [(x\pm\frac{%
\Theta}{2})^{2}+\frac{\Gamma^{2}}{4}]}= \frac{\pi(3\frac{\Theta}{2}\pm i%
\frac{3}{2}\Gamma)}{\Gamma(\frac{\Theta}{2}\pm i\frac{\Gamma}{2}) (\frac{%
\Theta}{2}^{2}+\frac{9}{4}\Gamma^{2})}
\end{eqnarray}
from eqs.(A1-A5), we can obtain obtain an exact expression of $\beta(T,t)$,
but we are only interested in the behavior of the short time, because after
evolution of the long time, all states of the system will return to vacumm
reduced by the eviornment. So in the condition of $\Gamma t\ll 1$ and $\frac{%
\Theta}{2} t \ll 1$, we only keep the first order small quantity, then we
have
\begin{equation}
\beta(T,t)=\bar{n} \left [ \frac{L\eta^{2}\Gamma (2\frac{\Theta}{2}%
^{2}+\Gamma^{2}) (\frac{\Theta}{2}^{2}+2\Gamma^{2})}{2c(\frac{\Theta}{2}^{2}
+\frac{\Gamma^{2}}{4})^{2}(\frac{\Theta}{2}^{2}+\frac{9}{4}\Gamma^{2})}
\right ]\Gamma t=\bar{n}{\cal D}\Gamma t
\end{equation}
which is needed in eq.(36).
\begin{multicols}{2}

\end{multicols}

\end{document}